\documentclass[aps,prl,twocolumn,,superscriptaddress,floatfix]{revtex4-1}
\usepackage[svgnames]{xcolor}
\usepackage[colorlinks=true,urlcolor = purple,linkcolor=teal,citecolor=magenta,bookmarks=false]{hyperref}
\usepackage{amsfonts,amsmath,amssymb}
\usepackage{graphicx}
\usepackage{dsfont}
\usepackage{graphicx}
\usepackage{amssymb}
\usepackage{amsmath}
\usepackage{empheq}
\usepackage{bm}
\usepackage{ulem}
\usepackage{stmaryrd}
\usepackage{tkz-graph}
\usepackage{tikz}
\usetikzlibrary{arrows,%
                petri,%
                topaths}%
\usepackage{tkz-berge}
\usetikzlibrary{matrix,shapes,arrows,decorations.markings}
\usepackage[nomessages]{fp}
\newcommand{\bc}{\begin{center}}
\newcommand{\ec}{\end{center}}
\newcommand{\beq}{\begin{equation}}
\newcommand{\eeq}{\end{equation}}
\newcommand{\beqa}{\begin{eqnarray}}
\newcommand{\eeqa}{\end{eqnarray}}
\newcommand{\beqs}{\begin{eqnarray*}}
\newcommand{\eeqs}{\end{eqnarray*}}
\newcommand{\bi}{\begin{itemize}}
\newcommand{\ei}{\end{itemize}}

\newcommand{\bra}{\langle}
\newcommand{\ket}{\rangle}

\newcommand{\Tr}{{\rm Tr}}
\newcommand{\Det}{{\rm Det}}
\newcommand{\id}{{\rm Id}}
\begin{document}

\title{\bf  Universal fluctuations around typicality for quantum ergodic systems}

\author{Michel Bauer}
\affiliation{Institut de Physique Th\'eorique de Saclay, CEA-Saclay $\&$ CNRS, 91191 Gif-sur-Yvette, France.}
\affiliation{D\'epartement de Math\'ematiques et Applications, ENS-Paris, 75005 Paris, France.}
\author{Denis Bernard}
\author{Tony Jin}
\affiliation{Laboratoire de Physique de l'Ecole Normale Sup\'erieure de Paris, CNRS, ENS $\&$ Universit\'e PSL, Sorbonne Universit\'e, Universit\'e de Paris, 75005 Paris, France}

\date{\today}

\begin{abstract} 
For a quantum system in a macroscopically large volume $V$, prepared in a pure state and subject to maximally noisy or ergodic unitary dynamics, the reduced density matrix of any sub-system $v\ll V$ is almost surely totally mixed. We show that the fluctuations around this limiting value, evaluated according to the invariant measure of these unitary flows, are captured by the Gaussian unitary ensemble (GUE) of random matrix theory. An extension of this statement, applicable when the unitary transformations conserve the energy but are maximally noisy or ergodic on any energy shell, allows to decipher the fluctuations around canonical typicality. According to typicality, if the large system is prepared in a generic pure state in a given energy shell, the reduced density matrix of the sub-system is almost surely the canonical Gibbs state of that sub-system. We show that the fluctuations around the Gibbs state are encoded in a deformation of the GUE whose covariance is specified by the Gibbs state. Contact with the eigenstate thermalisation hypothesis (ETH) is discussed. 
\end{abstract}

\maketitle

{\bf {Introduction}.--} 
Understanding how isolated quantum systems relax towards thermal equilibrium is a central question in quantum statistical physics. There are essentially two lines of thought in the literature: an ensemblist point of view, according to which a system in thermal equilibrium is represented by a statistical ensemble of states close to the canonical or micro-canonical mixed state, and a purist perspective, according to which a large system in a single pure state may behave as thermal for any local measurement. See \cite{Gibbs,Jancel13,Deutsch18} for thorough discussions. These understandings led to the related notions of typicality \cite{Tasaki16,Goldstein06,Winter06}, partially present in Von Neumann’s work \cite{ergodic}, and of eigen-state thermal hypothesis (ETH) \cite{Deutsch91,Srednicki94,Rigol08}, formalising under which conditions a pure state can describe thermal equilibrium locally.

Thermalisation occurs via entanglement which allows information to spread over the system. Recent studies of random quantum circuits \cite{Chamon1,Q-circuit1,Q-circuit2,Gullans19} gave detailed hints on this mechanism. Thermalisation may be quantified by looking at properties of the system state reduced to small sub-systems, such as the entanglement spectrum~\cite{Haldane}. It is now believed that there is a link between thermalisation and the distribution of the entanglement spectrum~\cite{P-WD-thermal1,P-WD-thermal2}, e.g. whether it is Poisson or Wigner-Dyson like. For instance, entanglement spectrum statistics have been used as diagnosis for thermalisation, distinguishing Clifford circuits from universal quantum circuits~\cite{Chamon19}. This establishes a link between thermalisation, which is a statement about typical behaviour, and fluctuations, which goes beyond typical behaviours.

This letter aims at making this link more precise, and possibly rigorous. Under the assumption of some form of ergodicity, we present a simple derivation and characterisation of the fluctuations of density matrices reduced to small sub-systems embedded in a large system prepared in a pure state. The form of these fluctuations makes contact with random matrix theory. It describes density matrix fluctuations beyond their almost sure thermal typical behaviours.

{\bf {Maximally noisy or ergodic quantum systems}.--}
Let us consider a quantum system in a large volume $V$ with Hilbert space $\mathcal{H}_V$, with $\mathrm{dim}\mathcal{H}_V\approx q^V$ with $q$ counting the number of local degrees of freedom. 
We assume unitary (non-dissipative, but possibly noisy) dynamics on $\mathcal{H}_V$, mapping the system density matrix $\rho$ into $U\rho\, U^\dag$ with $U\in  U(\mathcal{H}_V)$. Although the dynamics can be discrete or continuous in time, we shall focus on the discrete time setting to ease the description. The evolution of the system state is then implemented by iterative products,
\beq \label{eq:product}
 U_{p}\cdots U_{2}U_{1},
 \eeq
where the $U_{j}$'s, which may be random or not, code for the system evolution from time $t_j$ to $t_{j+1}$. In the noisy cases, they are randomly picked from a sampling set, so that the time evolved state $\rho$ is random. 

By assumption, unitary flows generated by \eqref{eq:product} will be declared to be maximally ergodic \cite{ergodic,Comment0}, or maximally noisy, if the iterated products \eqref{eq:product} cover densely the unitary group $U(\mathcal{H}_V)$. Examples are provided by (possibly random) quantum circuits based on a set of universal quantum gates \cite{Qinfo-book,randomU2,randomU3}. Time continuous maximally noisy dynamics on spin chains were recently analysed as quantum diffusive model systems \cite{Knap18,BBJ19,Knap18,Lama18,Gullans18}. 

We shall be interested in the long time steady distribution of the system state. This statistical distribution is encoded in the invariant measure of the flows \eqref{eq:product}, which we assume to be unique. Ergodicity can alternatively be defined by demanding this measure to be $U(\mathcal{H}_V)$-invariant \cite{Comment0}. If ergodicity (as defined above) holds, time averages of expectations coincide at large time with statistical averages measured with the invariant measure. The system being closed (except for possible exterior noise), we take it to be initially in a pure state $|\psi_0\ket$, so that the system initial density matrix $\rho_0=|\psi_0\ket\bra\psi_0|$ has rank one.

We consider a sub-system of volume $v\ll V$ of the original system, with Hilbert space $\mathcal{H}_v$, $\mathrm{dim} \mathcal{H}_{v}\approx q^v$, and look at the distribution of its reduced density matrix $\rho_v$ defined by tracing out the degrees of freedom of the complement $V\setminus v$ of $v$ in $V$: $\rho_v:=\Tr_{\mathrm{H}_{V\setminus v}}\rho$ with $\mathcal{H}_V=\mathcal{H}_v \otimes \mathcal{H}_{V\setminus v}$. The system $V\setminus v$ plays the role of sub-system environnement.
We claim that, in the large volume limit (with both $v$ and $V$ large, but $V\gg v$), the law of this reduced density matrix is
\begin{equation} \label{eq:claim}
\lim_{N\gg M \to \infty} \mathrm{law}\,\sqrt{N}\, \Big( \rho_v - \frac{\id_{\mathcal{H}_{v}}}{M} \Big) \equiv  \mathrm{GUE},
\end{equation}
where $\mathrm{GUE}$ refers to the Gaussian unitary ensemble of $n\times n$ hermitian matrices in the large $n$ limit \cite{RMT}. This statement is valid in the limit $N\to\infty$ first, then $M\to\infty$, with $N=\mathrm{dim} \mathcal{H}_V$ and $M=\mathrm{dim} \mathcal{H}_{v}$. Eq.\eqref{eq:claim} implies that, for $N\gg M\gg 1$, the reduced density matrix is almost totally mixed, $\rho_v\simeq \frac{\id_{\mathcal{H}_{v}}}{M}$, plus small Gaussian fluctuations of order $1/\sqrt{N}$ controlled by the GUE.

At finite but large $N$ and $M$, ($N\gg M\gg 1$), we have to keep track of sub-leading terms which do not directly contribute to the above limit \eqref{eq:claim}. As shown below (and in the Supplementary Material), we may approximate the statistics of the reduced density matrix by Gaussian matrices with mean and covariance,
\begin{eqnarray} \label{eq:finiteNMa}
\mathbb{E}_\mathrm{inv}[\Tr\, a\, \rho_v]^c &=& \frac{\Tr a}{M},\\
\mathbb{E}_\mathrm{inv}[(\Tr\, a\, \rho_v)^2]^c &=& \frac{1}{N}\Big( \frac{\Tr a^2}{M} -  (\frac{\Tr a}{M})^2\Big),
\label{eq:finiteNMb}
\end{eqnarray}
with $a$ any generic $M\times M$ matrix in $B(\mathcal{H}_v)$ and where the upper-script ${}^c$ refers to connected correlation functions. Higher cumulants of order $p$ decrease as $(\frac{1}{N})^{p-1}$ with $p\geq 3$.
These relations are clearly compatible with the constraint $\Tr \rho_v =1$. They were indirectly tested in \cite{P-WD-thermal1,P-WD-thermal2,Chamon19} by numerically computing the level statistics of the reduced density matrix in universal random quantum circuits initially prepared in pure factorised states.

Eqs.(\ref{eq:finiteNMa},\ref{eq:finiteNMb}) have to be understood in the large $N$ limit at $M$ fixed. If we pick a sub-system $v$ which is a fraction of the total volume, say $v=fV$, with $0<f<1$, both $M$ and $N$ become large but $M=N^f\ll N$. The scaling is then different, but the fluctuations are still GUE with $N^\frac{1+f}{2}\Big( \rho_v - \frac{\id_{\mathcal{H}_{v}}}{M} \Big)$ Gaussian in the large $N$ limit. See the Supplementary Material.

The previous results can be extended to deal with questions about foundations of quantum statistic physics.
From basics of thermodynamics, it is well know that the state of a system $v$, weakly coupled to a bath $B\equiv V\setminus v$, is described by the canonical Gibbs ensemble if the total system $V\equiv v + B$ is described by the micro-canonical ensemble. Canonical typicality \cite{Tasaki16,Goldstein06,Winter06} asserts a stronger statement: if the total system is prepared in a single pure state, not a mixture, in the eigen-space corresponding to the energy shell of the micro-canonical ensemble then, in the thermodynamical limit, the reduced density matrix of the sub-system $v$ is almost surely the Gibbs state.

Consider a small, but thermodynamically large, sub-system $v$ embedded in a large reservoir system $V$. Assume $v$ and $V\setminus v$ to be weakly coupled so that the total hamiltonian $H_V$ decomposes as $H_V = H_v + H_{V\setminus v} + H_\mathrm{int}$ with $H_\mathrm{int}$ negligible compared to $H_v$ or $H_{V\setminus v}$. Indeed, for short range interaction, the spectra of $H_V$, $H_{V\setminus v}$ or $H_v$, are expected to be extensive in their respective volume sizes, while that of the interaction $H_\mathrm{int}$ scales with the boundary area of $v$. Let $\rho_{\mathcal{E};v}$ be the reduced density matrix of $v$, assuming the total system $V$ be prepared in a pure state in the energy shell $\mathcal{E}$. As we shall argue below:  if the unitary dynamics on the total system conserves energy but is maximally noisy or ergodic in any energy shell then, in the thermodynamic limit, the fluctuations of the reduced density matrix around the typical Gibbs state are universal in the sense that
\beq \label{eq:GUEGibbs}
\lim_{V\gg v\to \infty} \mathrm{law} \sqrt{\Omega_V(\mathcal{E};\delta \mathcal{E})}\Big(\rho_{\mathcal{E};v} - \rho_\mathrm{Gibbs}\Big)
\equiv \mathrm{GUE}_\mathrm{Gibbs},
\eeq
where $\mathrm{GUE}_\mathrm{Gibbs}$ is the limit $v\to\infty$ of the law of Gaussian matrices in $\mathcal{H}_v$ with zero mean and covariance  $\mathbb{E}[(\Tr a\rho_{\mathcal{E};v})^2]= \Tr (a^2 \rho_\mathrm{Gibbs})$, where $\rho_\mathrm{Gibbs}$ is Gibbs state at temperature $1/\beta$ fixed by the reservoir energy $\mathcal{E}$. Here, $\Omega_V(\mathcal{E};\delta \mathcal{E})$ is the number of states of energy $\mathcal{E}$, up to $\delta \mathcal{E}$, in the total system $V$.

As above, this result is valid in the thermodynamic limit, first $V\to\infty$ then $v\to\infty$.  But a Gaussian approximation is valid at large but finite sub-system (with $V\gg v$ large) with the following mean and covariance:
\begin{eqnarray} \label{eq:finiteGibbs}
\mathbb{E}[\Tr\, a\rho_{\mathcal{E};v}]^c &=& \Tr (a \rho_\mathrm{{Gibbs}})\\
\mathbb{E}[(\Tr\, a\rho_{\mathcal{E};v})^2]^c &=& \frac{1}{\Omega_V(\mathcal{E};\delta \mathcal{E}) } \big(\Tr a^2\rho_\mathrm{Gibbs} -  (\Tr a\rho_\mathrm{Gibbs})^2\big) \nonumber
\end{eqnarray}
These fluctuations are of course exponentially small, scaling with the volume size as $1/{\sqrt{\Omega_V(\mathcal{E};\delta \mathcal{E})} }\approx e^{ -\frac{1}{2} S_V(\mathcal{E}) }$, with $S_V(\mathcal{E})$ the entropy of the total system. All higher cumulants are exponentially suppressed, by higher power of $e^{ -\frac{1}{2}S_V(\mathcal{E}) }$, in the thermodynamic limit. Nevertheless, they may probably be tested in mesoscopic quantum systems similarly as their companions were tested in random quantum circuits.

The statement \eqref{eq:finiteGibbs} bares similarities with consequences of the eigen-state thermalisation hypothesis (ETH)~\cite{Deutsch91,Srednicki94,Rigol08,Deutsch18}. The ETH asserts that matrix elements of local observables ${O}$, in the energy eigen-basis $|\mathcal{E}_i\rangle$, $H_V|\mathcal{E}_i\rangle=\mathcal{E}_i|\mathcal{E}_i\rangle$, take the following form: $O_{ij}=\mathcal{O}(\mathcal{E})\delta_{ij}+e^{-\frac{1}{2}S_V(\mathcal{E})} f_O(\mathcal{E},\omega)R^O_{ij}$, with $\mathcal{E}=\frac{1}{2}(\mathcal{E}_i+\mathcal{E}_j)$ and $\omega=\frac{1}{2}(\mathcal{E}_i-\mathcal{E}_j)$, with $\mathcal{O}$ and $f_O$ smooth functions, fastly decreasing with $\omega$, and $R^O_{ij}$ matrices of order one with erratically varying elements in the range of energy around $\mathcal{E}$, with zero mean. As in \eqref{eq:finiteGibbs}, higher cumulants of $R^O_{ij}$ are suppressed by higher powers of $e^{-\frac{1}{2}S_V(\mathcal{E})}$~\cite{Foini18}. Statistical unitary invariance of $R^O_{ij}$ has been tested in \cite{FoiniKurchan19}. The main point of ETH is to not introduce a priori a statistical ensemble but let it emerge from the fluctuations of the matrix elements $R^O_{ij}$ in the micro-canonical energy window. Sub-leading fluctuations to the thermodynamic expectations arise from the arbitrariness in picking a state in the energy shell $\mathcal{E}$ up to $\delta\mathcal{E}$. Ergodicity, as used above, replaces this arbitrariness by a measure symmetric under unitary transformations within this energy shell. Following typicality claims \cite{Tasaki16,Goldstein06,Winter06}, it indicates that ETH holds almost surely (w.r.t. to this uniform measure) for states in the energy shell $\mathcal{E}$. Of course, there could exist rare states, of measure zero, for which it does not hold \cite{scars}.
It is tempting to conjecture that, if the ETH holds for all local observables, then the invariant measure of the system dynamics, restricted to any energy shell, is unique.

{{\bf Reduced density matrix fluctuations}.--}
Let us now present the arguments leading to \eqref{eq:claim} and \eqref{eq:GUEGibbs}. The proof consists in first determining (exactly) the generating function of the total system state distribution, and second, in analysing the reduction of this generating function to the sub-system density matrix.

Let $\mathbb{E}_\mathrm{inv}$ be the measure on density matrices on $\mathcal{H}_V$ invariant under the flows \eqref{eq:product}. It depends on the initial state $\rho_0$. Given $\rho_0$, up to rotation, we assume unicity of  $\mathbb{E}_\mathrm{inv}$~\cite{Comments2}.  By construction, it is $U(\mathcal{H}_V)$-invariant for maximally noisy or ergodic dynamics, since the iterated products \eqref{eq:product} explore uniformly the unitary group $U(\mathcal{H}_V)$. 
This invariance, plus the fact that unitary transformations are spectrum preserving, determine the measure entirely. Its generating function $Z[A]$, 
\beq \label{eq:ZA}
 Z[A]:= \mathbb{E}_\mathrm{inv}\big[ e^{\Tr(A\rho)}\big], 
 \eeq
with $A\in B(\mathcal{H}_V)$, is represented by the Harish-Chandra-Itzykson-Zuber (HCIZ) integral \cite{HCIZ} over $U(\mathcal{H}_V)$, known in random matrix theory~\cite{RMT}. Namely
\beq \label{eq:HCIZ}
 Z[A] = \int_{U(\mathcal{H}_V)} \hskip -0.5 truecm dV\, e^{\Tr(V^\dag AV\rho_0)} .
 \eeq
 with $\rho_0$ the initial state. For $\rho_0$ a pure state, this distribution is also known as that of `Haar states' \cite{Page}. Entanglement spectrum of reduced density matrices of Haar states has been studied in \cite{Majumdar10,Hamma12,Ludwig17} by using Wishart ensembles \cite{Wishart,Lloyd88} (See the Supplementary Material for a brief discussion).

Indeed, the $U(\mathcal{H}_V)$-invariance means that, for any unitary $V$, the density matrices $\rho$ and $V\rho V^\dag$ are identically distributed. As a consequence, the generating function $Z[A]$ is $U(\mathcal{H}_V)$-invariant: $Z[V^\dag AV]=Z[A]$ for any $V\in  U(\mathcal{H}_V)$.
Hence, by integrating over the unitary group with the normalised Haar measure, 
\beq
 Z[A] =  \int_{U(\mathcal{H}_V)}  \hskip -0.5 truecm dV\, \mathbb{E}_\mathrm{inv}[e^{\Tr(V^\dag AV\rho)} ].
\eeq
Permuting integration and expectation, we write $Z[A] =  \mathbb{E}_\mathrm{inv}[\int_{U(\mathcal{H}_V)} dV\, e^{\Tr(V^\dag AV\rho)} ]$. By invariance of the Haar measure, the function $\int_{U(\mathcal{H}_V)} dV\, e^{\Tr(V^\dag AV\rho)}$ to be averaged depends only the spectrum of $\rho$. Since the flows \eqref{eq:product} are unitary, they preserve this spectrum which is hence non-random. So, we can replace the random variable $\rho$ by the non-random initial data $\rho_0$, and pull out the integral from the expectation. This proves \eqref{eq:HCIZ}. 

The HCIZ integral \eqref{eq:HCIZ} can be computed exactly when $\rho_0$ is of rank one, with output 
 \beq
\sum_{n\geq 0}\frac{(N-1)!}{(N+n-1)!} \sum_{n_k,\, \sum_{k\geq 1} kn_k=n} \prod_k \frac{1}{n_k!}\left(\frac{\Tr A^k}{k}\right)^{n_k} , 
\eeq
with $N=\mathrm{dim}\mathcal{H}_V$. For $N\gg 1$, we may naively approximate $\frac{(N-1)!}{(N+n-1)!}$ by $N^{-n}$ and sum the series to get
 \begin{equation} \label{eq:Z-form}
 Z[ A ] = \frac{1}{\Det(\id_{\mathcal{H}_V}- A/N)} .
 \end{equation}
This relation has to be understood in the sense that $Z[NA]=\Det(\id_{\mathcal{H}_V}- A)^{-1}$ in the limit $N\to\infty$. 
The asymptotic expansion of $Z[A]$ in powers of $\frac{1}{N}$ contains  sub-leading corrections that might have to be taken into account. To simplify the presentation, the following arguments are going to be based on \eqref{eq:Z-form}, and we refer to the Supplementary Material for a more detailed proof.

Consider now a sub-system of volume $v\ll V$ of the original system. With local degrees of freedom, the original Hilbert space factorises as $\mathcal{H}_V=\mathcal{H}_v\otimes \mathcal{H}_{V\setminus v}$ with $\mathcal{H}_v$ (resp. $\mathcal{H}_{V \setminus v}$) the Hibert space of the sub-system (resp. of its complement). Let $\rho_v=\Tr_{\mathcal{H}_{V\setminus v}}\rho$ be the sub-system reduced density matrix. Since $\Tr(a\otimes \id_{\mathcal{H}_{V\setminus v}}\rho) = \Tr\, a \rho_v$, its generating function $Z_v[a]$ is given by $Z[A]$ but with $A=a\otimes \id_{\mathcal{H}_{V\setminus v}}$.  From \eqref{eq:Z-form}, it then follows that (for $\rho_0$ a pure state):
\begin{equation}
Z_v[a]:=  \mathbb{E}_\mathrm{inv}\big[ e^{\Tr(a\rho_v)}\big] = \Big(\frac{1}{\Det(\id_{\mathcal{H}_{v}}- a/N)}\Big)^{N/M} ,
\end{equation}
with $M:=\mathrm{dim} \mathcal{H}_{v}$ and $\mathrm{dim} \mathcal{H}_{V\setminus v}=N/M$.

Let us now expand $W_v[a]:=\log Z_v[a]$ in power of $a$. The first term is $\frac{\Tr(a)}{M}$, which implies that $\rho_v$ is totally mixed in the large system size limit: $\rho_v\simeq \frac{\id_{\mathcal{H}_{v}}}{M}$. The second term in $W_v[a]$ reads $\frac{\Tr(a^2)}{2MN}$. So, we rescale $a\to\sqrt{NM}a$, set $\rho_v= \frac{\id_{\mathcal{H}_{v}}}{M}+ \delta\rho_v$, and consider
\begin{equation}
\mathbb{E}_\mathrm{inv}\big[ e^{\sqrt{NM} \Tr(a\delta \rho_v)}\big] =\frac{ e^{- \sqrt{\frac{N}{M}}\,\Tr(a)} }{ \Det\big(\id_{\mathcal{H}_{v}}- \sqrt{\frac{M}{N}} a\big)^{N/M} }.
\end{equation}
We already checked that, when expanding in power of $a$, the first term vanishes and the second is appropriately normalised. The higher order term at order $k$ is $\frac{1}{k} \Tr a^k \, (M/N)^{k/2}(N/M)$. It scales as $(\frac{M}{N})^{k/2-1}$ and hence vanishes for $k>2$ in the large system size $N\gg M$. We thus proved that the generating function is Gaussian, 
\begin{equation} \label{eq:main}
\lim_{N\gg M \to \infty} \mathbb{E}_\mathrm{inv}\big[ e^{\sqrt{NM} \Tr(a\delta \rho_v)}\big] = e^{\frac{1}{2} \Tr a^2 },\
\end{equation}
order by order in $a$, in the large system size. 

Eq.\eqref{eq:main} means that
\begin{equation}
\rho_v = \frac{\id_{\mathcal{H}_{v}}}{M} + \frac{1}{\sqrt{NM}}\, R_v +\cdots,
\end{equation}
with $R_v$ a Gaussian hermitian matrix with variance one. Alternatively, since GUE matrices scale as $1/\sqrt{M}$ in the usual normalisation, this proves eq.\eqref{eq:claim}. At finite but large $N$ and $M$, with $N\gg M$, we have to keep track of sub-leading terms that we neglected above in \eqref{eq:Z-form}. As shown in the Supplementary Material, we may then approximate the statistics of the reduced density matrix as a Gaussian matrix with mean and covariance (\ref{eq:finiteNMa},\ref{eq:finiteNMb}).

{\bf {Fluctuations around typicality}.--} 
Assume now that the unitary flows commute with a hamiltonian defined on $\mathcal{H}_V$ so that the energy is a conserved quantity~\cite{Comment3}. Let $\mathcal{H}_{V;\mathcal{E}}$ be the Hilbert space spanned by all eigenstates of energy less than $\mathcal{E}$, so that $\delta\mathcal{H}_{V;\mathcal{E}}$ is the Hilbert space of states with energy between $\mathcal{E}$ and $\mathcal{E}+\delta\mathcal{E}$.  Let $\Omega_V(\mathcal{E}):=\mathrm{dim}\,\mathcal{H}_{V;\mathcal{E}}$ be the number of eigenstate in $V$ of energy less that $\mathcal{E}$. We set $\Omega_V(\mathcal{E};\delta\mathcal{E}):=\mathrm{dim}\,\delta\mathcal{H}_{V;\mathcal{E}}$. 

We start by assuming that the total system has been prepared in a pure state $\rho_0$ with energy between $\mathcal{E}$ and $\mathcal{E}+\delta\mathcal{E}$. Since the unitary transformations commute with the hamiltonian, they induce a flow on $\delta\mathcal{H}_{V;\mathcal{E}}$. We assume this flow to be maximally noisy or ergodic. Let $\rho_\mathcal{E}$ be the state on $\delta\mathcal{H}_{V;\mathcal{E}}$ generated by this flow. By the same reasoning as previously, the invariant measure is $U(\delta\mathcal{H}_{V;\mathcal{E}})$-invariant~\cite{Comment-M}. Hence, its generating function is given by the HCIZ integral but on the unitary group $U(\delta\mathcal{H}_{V;\mathcal{E}})$ and, for $\rho_0$ a pure state, it reads
\begin{equation}
 Z_\mathcal{E}[ A ]:= \mathbb{E}_\mathrm{inv}[e^{\Tr(A\rho_\mathcal{E})}] = \frac{1}{\Det(\id_{\delta\mathcal{H}_{V;\mathcal{E}}}- {A}/{\Omega_V(\mathcal{E};\delta\mathcal{E})})},
 \end{equation}
 with $A\in B(\delta\mathcal{H}_{V;\mathcal{E}})$. The main difference with the previous discussion is that all objects ($\rho_\mathcal{E}$, $A$, $\Tr(\cdot)$, $\Det(\cdot)$, $\cdots$) are now restricted to the subspace $\delta\mathcal{H}_{V;\mathcal{E}}$ of energy $\mathcal{E}$ up to $\delta\mathcal{E}$.

Again, we consider a sub-system of volume $v$ in $V$, and we view the complement $V\setminus v$ as a bath environnement for $v$. 
Let $\rho_{\mathcal{E};v}$ be the density matrix reduced to $v$. Let $Z_{\mathcal{E};v}[a]:=\mathbb{E}[e^{\Tr(a\rho_{\mathcal{E};v})}]$ be the generating function of its distribution and $W_{\mathcal{E};v}[a]:=\log Z_{\mathcal{E};v}[a]$ the generating function of its cumulants. 

We need to recall the connection between canonical and micro-canonical ensembles, from basic thermodynamics. Neglecting, as usual, the interaction energy, which is proportional to the area of the boundary of $v$, we can decompose the total energy $\mathcal{E}$ as the sum of the energy $E$ in the sub-system $v$ and the energy $\mathcal{E}-E$ in the complement $V\setminus v$. As a consequence, we have an (approximate) decomposition of $\delta\mathcal{H}_{V;\mathcal{E}}$ as a direct sum of eigen-spaces of the respective hamiltonians in the sub-system $v$ and its complement. This translates into the following resolution of the identity on $\delta\mathcal{H}_{V;\mathcal{E}}$
\beq \label{eq:resolution}
 \id_{\delta\mathcal{H}_{V;\mathcal{E}}}= \int_E dP_{v;E}\otimes  \id_{\delta\mathcal{H}_{V\setminus v;\mathcal{E}-E}} ,
\eeq
where $dP_{v;E}$ is the spectral projector on eigenvalue between $E$ and $E+dE$ for the hamiltonian on the sub-system $v$. Let $\Omega_v(E):=\mathrm{rank}\,dP_{v;E}$ be the number of eigen-state of energy $E$, up to $dE$, in the sub-system $v$. By Boltzmann relation, we have $\log \Omega_V(\mathcal{E};\delta\mathcal{E}) \approx S_V(\mathcal{E})$ with $S_V(\mathcal{E})$ the micro-canonical entropy of the total system. For $v\ll V$, and by extensitivity of the spectrum $E\ll\mathcal{E}$, so that $S_{V\setminus v}(\mathcal{E}-E)=S_V(\mathcal{E})- \beta E+\cdots$ where $\beta:= dS_V(\mathcal{E})/d\mathcal{E}$. Hence, $\Omega_{V\setminus v}(\mathcal{E}-E;\delta \mathcal{E})\approx e^{-\beta E}\, \Omega_{V}(\mathcal{E};\delta \mathcal{E})$, up to a multiplicative constant. Using the sum rule $\Omega_V(\mathcal{E};\delta \mathcal{E})=\int_E d\Omega_v({E}) \Omega_{V\setminus v}(\mathcal{E}-E;\delta\mathcal{E})$, which follows from \eqref{eq:resolution}, this constant is found to be the sub-system partition function,  $Z_v= \int_E d\Omega_v({E})\, e^{-\beta E}$. Thus, the ratio $\frac{\Omega_{V\setminus v}(\mathcal{E}-E;\delta\mathcal{E})}{\Omega_V(\mathcal{E};\delta \mathcal{E})}$ is equal to the Boltzmann weight at temperature $1/\beta$ specified by the total system,
\beq \label{eq:boltzmann}
\frac{\Omega_{V\setminus v}(\mathcal{E}-E;\delta\mathcal{E})}{\Omega_V(\mathcal{E};\delta \mathcal{E})} = \frac{e^{-\beta E}}{Z_v} .
\eeq
This had to be expected as the ratio $\frac{\Omega_{V\setminus v}(\mathcal{E}-E;\delta\mathcal{E})}{\Omega_V(\mathcal{E};\delta \mathcal{E})}$ is the probability in the micro-canonical ensemble for the sub-system $v$ to be in a state of energy $E$.

Let us now look at the first term in $W_{\mathcal{E};v}[a]$. It reads $\Tr \frac{a\otimes \id}{\Omega_V(\mathcal{E};\delta \mathcal{E})}$.
Inserting the resolution of the identity \eqref{eq:resolution}, we get
\[ 
\int_E \Tr (dP_{v;E}\, a) \, \frac{\Omega_{V\setminus v}(\mathcal{E}-E;\delta\mathcal{E})}{\Omega_V(\mathcal{E};\delta \mathcal{E})}.
\]
Using \eqref{eq:boltzmann}, this reads,
\beq
 \int_E \Tr (dP_{v;E}\frac{e^{-\beta E}}{Z_v}\, a) = \Tr (\rho_\mathrm{Gibbs}\, a),
 \eeq
with, as it should be, $\rho_\mathrm{Gibbs}=\int_E dP_{v;E}\frac{e^{-\beta E}}{Z_v}$, the thermal Gibbs state at temperature $1/\beta$.

The second term in $W_{\mathcal{E};v}[a]$ reads $\frac{1}{2} \Tr \big(\frac{a\otimes \id}{\Omega_V(\mathcal{E};\delta \mathcal{E})}\big)^2$. Inserting again the resolution of identity \eqref{eq:resolution}, we get
\[ 
\frac{1}{2}\int_E \Tr (dP_{v;E}\, a^2)\, \frac{\Omega_{V\setminus v}(\mathcal{E}-E;\delta\mathcal{E})}{\Omega_V(\mathcal{E};\delta \mathcal{E})^2},
\]
which reads, using again \eqref{eq:boltzmann},
\beq \frac{1}{2\Omega_V(\mathcal{E};\delta \mathcal{E})} \Tr_{\mathcal{H}_E} (a^2 \rho_\mathrm{Gibbs}) .
\eeq
By power counting, all higher order terms are suppressed by higher powers of ${1}/{\Omega_V(\mathcal{E};\delta \mathcal{E})}$. 
This proves the claim \eqref{eq:GUEGibbs}. This result is valid in the large volume limit, but we can use it to give a Gaussian approximation of the reduced density matrix statistics valid at large but finite volume (with $V\gg v$), with mean and covariance \eqref{eq:finiteGibbs}.

{\bf {Conclusion}.--} 
We have provided simple arguments, if not proofs, yielding to a characterisation of the leading fluctuations of the reduced density matrices of sub-systems embedded in a thermodynamically large closed environnement. These arguments are grounded in the symmetries of the invariant measure of the underlying unitary dynamics, reflecting their assumed ergodicity. We pointed to a possible link between the ETH and the uniqueness of the invariant measure,

Extending these arguments to non-equilibrium situations, and determining the fluctuations of density matrices reduced to small cells within a large volume, will yield a better understanding of the emergence of fluctuating hydrodynamics, of its nature and its domain of validity, in extended quantum many-body systems out-of-equilibrium.

{\bf {Acknowledgments}.--} 
D.B. thanks Claudio Chamon for discussions.
We acknowledge support from the CNRS and the ANR via the contract ANR-14-CE25-0003.


\begin{thebibliography}{xxx}
\bibitem{Comment0} There is some discussions about what is the (or an) appropriate definition of quantum ergodicity. See eg. \cite{Deutsch91,Srednicki94,Deutsch18,ergodic} for a discussion.  At least for the purpose of this paper, we choose as definition the fact that the generators of the time evolution cover uniformly the unitary group, or alternatively, that the invariant measure is $U(\mathcal{H})$-invariant.
\bibitem{Comments2} Unicity can be proved in specific cases (e.g. see \cite{BBJ19}). It is guaranteed in ergodic systems, by definition.
\bibitem{Comment3} The construction we describe can be generalised to unitary flows preserving any number of conserved quantities.
\bibitem{Comment-M} This is a subtle point: a unitary flow commuting with the hamiltonian acts on each leaf of fixed energy separately, and hence cannot be ergodic on $\delta\mathcal{H}_{V;\mathcal{E}}$ for a fixed $\delta \mathcal{E}$. However, the ergodicity may be recovered in the limit $\delta\mathcal{E}\to 0$, but implementing this procedure requires commuting several limits and is mathematically delicate.
\end{thebibliography}

\newpage
~ \vfill \eject

\onecolumngrid 

\centerline{\Large {\bf Supplementary Material}}
\bigskip

\centerline{\bf  Universal fluctuations of reduced density matrices}
 \centerline{\bf  in maximally noisy or ergodic quantum systems and typicality}
\vskip 1.0 truecm

{\bf A- Harish-Chandra-Itzykson-Zuber integrals for a rank one matrix}

Recall the Harish-Chandra-Ittzykson-Zuber (HCIZ) integral over the unitary group $U(N)$:
\beq
 Z[A,B] = \int_{U(N)} dV\, e^{\Tr(VAV^\dag B)} .
 \eeq
 
Suppose that $\mathrm{rank} B=1$, i.e. $B$ is a projector.
We have two equivalent exact expressions:
\begin{eqnarray} \label{eq:HCIZ-N}
 Z_\mathrm{rank}[A]&=& \sum_n \frac{(N-1)!}{(N+n-1)!} \sum_{n_k,\, \sum_{k\geq 1} kn_k=n} \prod_k \frac{1}{n_k!}\left(\frac{\Tr A^k}{k}\right)^{n_k} \\
 &=& \sum_n \frac{(N-1)!}{(N+n-1)!}\, \chi_n(A),
\end{eqnarray}
with $\chi_n(A)$ the character of the rank $n$ totally symmetric representation of $SU(N)$. A proof of the first expression was given in \cite{BBJ19}, but it may probably be found elsewhere in the mathematical literature. The second formula follows from the character expansion of the HCIZ integral $Z[A,B]=\sum_Y c_Y\, \chi_Y(A)\chi_Y(B)$ where $ \chi_Y(A)$ are the $SU(N)$-characters in the representation labeled by the Young diagram $Y$ and $c_Y$ some known coefficients. If $B$ is of rank one, $\chi_Y(B)$ is non vanishing only for $Y$ the rank $n$ totally symmetric representation and in those cases $\chi_Y(B)=1$. Thus, for $B$ of rank one $Z_\mathrm{rank}[A]=\sum_n c_n\, \chi_n(A)$. The coefficients $c_n$ are found to be $c_n=1/n!d_n$, with $d_n$ the dimension of the rank $n$ totally symmetric representation, by demanding that $Z_\mathrm{rank}=e^t$ for $A=t\id$. The two expressions coincide (as they should) thanks to the relation between the character of the rank $n$ totally symmetric representation and the monomials $\prod_k  (\Tr A^k)^{n_k}$:
\beq \label{eq:charac_n}
\chi_n(A) =  \sum_{n_k,\, \sum_{k} kn_k=n} \prod_k \frac{1}{n_k!}\left(\frac{\Tr A^k}{k}\right)^{n_k} .
\eeq
This later relation is a direct consequence of the Cauchy identity
\[ \prod_{i,j} \left(\frac{1}{1-a_ib_j}\right) = \sum_Y  \chi_Y(A)\chi_Y(B),\]
with $A$ (resp. $B$) the diagonal matrices with entries $a_i$ (resp. $b_j$), specialised to the case where $B$ is a matrix of rank one. For $B$ of rank one, it gives $\prod_{i} \left(\frac{1}{1-ta_i}\right) = \sum_n  t^n\chi_n(A)$, which yields \eqref{eq:charac_n}.
\bigskip

{\bf B- Scaling limits}

Let us look at the large $N$ limit of \eqref{eq:HCIZ-N}. Approximating $\frac{(N-1)!}{(N+n-1)!} $ by $\frac{1}{N^n}$, we get, 
\begin{eqnarray}
 Z_\mathrm{rank}[A] \simeq  \sum_n \sum_{n_k,\, \sum_{k} kn_k=n} \prod_k  \frac{1}{n_k!}\left(\frac{\hat t_k}{kN^k}\right)^{n_k} = \exp\Big( \sum_k \frac{\hat t_k}{kN^k} \Big) ,
\end{eqnarray}
with $\hat t_k=\Tr A^k$, so that,
\beq \label{eq:Nrho-A}
Z_\mathrm{rank}[NA] \simeq  \exp\big( - \Tr \log (1 - A) \big) = \frac{1}{\Det(1-A)},
\eeq
at large $N$, under appropriate conditions on $A$.

Let us now be more careful by keeping track of the sub-leading $1/N$ factor. They play a role in ensuring that the distribution is compatible with the constraint $\Tr \rho =1$.

Let us expand $W_\mathrm{rank}[A]:=\log Z_\mathrm{rank}[A]$, keeping track of the sub-leading factors of $1/N$. For $\log Z_\mathrm{rank}[A]=1 + z_1 + z_2 + z_3 + \cdots$ with $z_k$ of degree $k$ in $A$, we have
\[ W_\mathrm{rank}[A] = z_1 + (z_2-\frac{1}{2}z_1^2) + (z_3 - z_1z_2 +\frac{1}{3}z_1^3) + \cdots .\]
We have $z_1= \tau_1$, $z_2=\frac{N}{N+1}(\tau_2 +\frac{1}{2}\tau_1^2)$, and $z_3=\frac{N^2}{(N+1)(N+2)}(\tau_3+\tau_1\tau_2+ \frac{1}{6}\tau_1^3)$ , with $\tau_k = \frac{\Tr A^k}{kN^k}$.\\
Hence, with $t_k=\frac{\Tr A^k}{N^k}=k\tau_k$,
\begin{eqnarray}
W_\mathrm{rank}[A] &=& t_1 + \frac{1}{2(N+1)}\big( Nt_2 -  t_1^2 \big) + 
\frac{1}{3(N+1)(N+2)}\big( N^2 t_3 - 3N t_1t_2 + 2t_1^3\big) +\cdots \\
&=& (\frac{\Tr A}{N}) + \frac{1}{2(N+1)}\Big(  (\frac{\Tr A^2}{N}) -  (\frac{\Tr A}{N})^2\Big)
+ \frac{1}{3(N+1)(N+2)}\Big( (\frac{\Tr A^3}{N}) - 3 (\frac{\Tr A}{N})(\frac{\Tr A^2}{N}) + 2 (\frac{\Tr A}{N})^3 \Big) +\cdots \nonumber 
\end{eqnarray}
We can check that $W_\mathrm{rank}[A=1]= 1$, as it should be (recall that $t_k= N^{1-k}$ for $A=1$)

For the reduced density matrix,  the generation function is (taking into account that $A=a\otimes 1$)
\beq 
W_\mathrm{red}[a] =  W_\mathrm{rank}[t_k= \frac{\Tr a^k}{MN^{k-1}}]
\eeq
From this we read the first few connected cumulants
\begin{eqnarray*}
\mathbb{E}[\Tr\, a\rho_v]^c &=& \frac{\Tr a}{M} ,\\
\mathbb{E}[(\Tr\, a\rho_v)^2]^c &=& \frac{1}{(N+1)}\Big( (\frac{\Tr a^2}{M}) -  (\frac{\Tr a}{M})^2\Big) ,\\
\mathbb{E}[(\Tr\, a\rho_v)^3]^c &=& \frac{2}{(N+1)(N+2)}\Big( (\frac{\Tr a^3}{M}) -3 (\frac{\Tr a^2}{M})(\frac{\Tr a}{M}) + 2(\frac{\Tr a}{M})^3 \Big).
\end{eqnarray*}
This is again compatible with $\Tr\, \rho_v =1$, as it should be.

We can now look at the large $N$ limit at $M$ fixed and prove that the fluctuations, properly rescaled, are Gaussian, as can be seen from the first few cumulants given above. To take into account the fact that fluctuations scale as $\sqrt{N}$, we start with $Z_\mathrm{rank}$ evaluated at $A(b)=b\otimes \id$ with $b=\sqrt{N}(a- \frac{\Tr a}{M})$. Then,
\beq 
Z_\mathrm{rank}[A(b)]= \mathbb{E}[ e^{ \sqrt{N}\, \Tr a(\rho_v- \frac{\id}{M})}],
\eeq
because $\Tr (a- \frac{\Tr a}{M})\rho_v =\Tr\, a(\rho_v- \frac{\id}{M})$ since $\Tr\rho_v=1$. We have $\Tr A(b)^k = \sqrt{N}^k\frac{N}{M}\Tr (a- \frac{\Tr a}{M})^k$ where the extra factor $(\frac{N}{M})$ comes from tracing over the identity in the complementary space of dimension $\frac{N}{M}$. Let write $\Tr A(b)^k = N^\frac{k+2}{2}\, \theta_k$ with $\theta_k = \frac{1}{M} \Tr (a- \frac{\Tr a}{M})^k$ independent of $N$. Note that $\theta_1=0$, by construction,  and $\theta_2= \frac{1}{M} \Tr (a- \frac{\Tr a}{M})^2=\frac{\Tr a^2}{M} -(\frac{\Tr a}{M})^2$. By \eqref{eq:HCIZ-N}, we then have
\begin{eqnarray}
Z_\mathrm{rank}[A(b)] &=& \sum_n \frac{(N-1)!}{(N+n-1)!} \sum_{n_k,\, \sum_{k} kn_k=n} \prod_k \frac{1}{n_k!}\left(\frac{N^\frac{k+2}{2}\, \theta_k}{k}\right)^{n_k} \\
&=& \sum_n \frac{(N-1)!N^n}{(N+n-1)!} \sum_{n_k,\, \sum_{k} kn_k=n} \left(\frac{1}{N}\right)^{\sum_k \frac{k-2}{2}n_k} \cdot \prod_k \frac{1}{n_k!}\left(\frac{\theta_k}{k}\right)^{n_k},
\end{eqnarray}
where we have pulled out the $N$-dependence. The pre-factors  $\frac{(N-1)!N^n}{(N+n-1)!} $ are of order $1+O(\frac{1}{N})$ at large $N$. Since $\theta_1=0$, $n_1$ has to be zero for the corresponding term in the above sum not to vanish. The factor $\left(\frac{1}{N}\right)^{\sum_k \frac{k-2}{2}n_k}$ is then of order one if and only if all $n_k=0$, for $k\geq 3$. Thus, the only terms which contribute at leading order $(\frac{1}{N})^0$ are those for which only $n_2\not =0$ and, we have, order by order in power of $a$, 
\begin{eqnarray}
Z_\mathrm{rank}[A(b)] &=& \sum_m \frac{1}{m!} \left(\frac{\theta_2}{2}\right)^{m}\, \left(1 + O(\frac{1}{N})\right) \\
&=&\exp{\frac{\theta_2}{2}} = \exp\left[{\frac{1}{2M} \Tr (a- \frac{\Tr a}{M})^2 }\right]
 \label{eq:Zgauss}
\end{eqnarray}
Because $ \frac{(N-1)!N^n}{(N+n-1)!}\left(\frac{1}{N}\right)^{\sum_{k\geq 2} \frac{k-2}{2}n_k}$, with $n= \sum_{k} kn_k$,  is  smaller than $1$ (for all $N$ and $n_k$), one can prove, using a dominated convergence argument, that this result is not only valid order by order in $a$ but uniformly in $(a- \frac{\Tr a}{M})$ in a neighbourhood of the origin.

Eq.\eqref{eq:Zgauss} proves that $X_M\equiv \sqrt{N}\, (\rho_v- \frac{\id}{M})$ is Gaussian, in the limit $N\to\infty$ at fixed $M$. The probability distribution of $X_M$ is that of a Gaussian random matrix with the constraint that $\Tr X_M =0$, that is
\beq
\mathcal{Z}^{-1}_M \int dX_M\, \delta(\Tr X_M)\, e^{-\frac{M}{2}\Tr X_M^2},
\eeq
where the integration is over the $M\times M$ hermitian matrices and $\mathcal{Z}_M$ a constant normalising the probability.

From this computation, we learn that, at finite but large $N$ and $M$, with $M\ll N$, we may approximate the statistic of the reduced density matrix as being Gaussian with mean and covariance:
\begin{eqnarray}
\mathbb{E}[\Tr\, a\rho_v]^c &=& \frac{\Tr a}{M} ,\\
\mathbb{E}[(\Tr\, a\rho_v)^2]^c &=& \frac{1}{N}\Big( \frac{\Tr a^2}{M} -  (\frac{\Tr a}{M})^2\Big) .
\end{eqnarray}

The scaling relation is different if $M$ scales with $N$, say $M=N^f$, with $0<f<1$, so that both $M$ and $N$ are large, still with $M\ll N$. It is easy to see that the correct scaling amounts to replace $\sqrt{N}$ by $N^\frac{1+f}{2}$ so that the Gaussian variable is
\beq
X_N := N^\frac{1+f}{2}\, \big(\rho_v- \frac{\id}{M}\big).
\eeq
Indeed, following the same strategy as above, we introduce $A_f:=N^\alpha(a- \frac{\Tr a}{M})\otimes \id$ with some exponent $\alpha$ to be fixed. Then $\Tr A_f^k = N^{k\alpha}\, \frac{N}{M} \Tr(a- \frac{\Tr a}{M})^k$ which scale as $N^{k\alpha+(1-f)}$. To apply the same argument as before, we have to compare $\prod_k (N^{k\alpha+(1+f)})^{n_k}$ to $(\frac{1}{N})^n$ with $n=\sum_k kn_k$. That is, we have to look at $\prod_k (N^{k(\alpha-1)+(1-f)})^{n_k}$ (with $k\geq 2$). For the scaling to be correct, we should have $2(\alpha-1)=f-1$, corresponding to $k=2$, and $k(\alpha-1)+(1-f)<0$ for $k\geq 3$. This fixes $\alpha =\frac{1+f}{2}$. The rest of the proof is then as before.
\bigskip

{\bf C- Singular case}

The story is different when $M$ and $N$ are both large but comparable $M\sim N$. Let $M$ be a fraction of $N$, that is $M= N/\ell$ with $\ell>1$ integer (dividing $N$).  Let us look at $ Z_\mathrm{rank}$ but with $A_N(a)=N\, a\otimes \id$. On one hand, we have
\beq
Z_\mathrm{rank}[A_N(a)] = \mathbb{E}[ e^{ {N}\, \Tr a\rho_v}]
\eeq
On the other hand, by \eqref{eq:HCIZ-N} with $\Tr A_N(a)^k = {N}^k\frac{N}{M}\Tr a^k= \ell N^k \Tr a^k$, we have
\begin{eqnarray}
Z_\mathrm{rank}[A_N(a)] &=&  \sum_n \frac{(N-1)!N^n}{(N+n-1)!} \sum_{n_k,\, \sum_{k} kn_k=n} \prod_k \frac{1}{n_k!}\left(\frac{\ell\Tr a^k}{k}\right)^{n_k} ,
\end{eqnarray}
where, again, we pull out the $N$ dependence using $\sum_{k} kn_k=n$. The term $ \frac{(N-1)!N^n}{(N+n-1)!} $ is of order $1$ for $N$ large. Thus, we get $ Z_\mathrm{rank}[A_N(a)] =\sum_{n_k} \prod_k \frac{1}{n_k!}\left(\frac{\ell\Tr a^k}{k}\right)^{n_k} + O(\frac{1}{N})$, so that
\begin{eqnarray}
\mathbb{E}[ e^{ {N}\, \Tr a\rho_v}]  =\exp \sum_k \frac{\ell\Tr a^k}{k}  =\left(\frac{1}{\Det(1- a)}\right)^{\ell},
\end{eqnarray}
for $N\gg1$ with $\ell=N/M$ fixed.
\bigskip 

{\bf D- Relation to Wishart ensembles}

The aim of the following is to make simple remarks on the relation between the previous results and Wishart ensembles of random matrices. These ensembles describe the statistical property of positive square matrices $\Omega:=Z Z^\dag$, with $Z$ complex rectangular matrices of different sizes, by fixing a measure on the $Y$'s, say
\beq \label{eq:wishart}
 dZdZ^\dag\, e^{-\Tr(Z Z^\dag)} ,
 \eeq
up to a multiplicative factor. Here, we connect these distributions with the Haar measure we used, say in the Harish-Chandra-Itzykson-Zuber integral \eqref{eq:HCIZ}. We start with the case of Haar pure states which, in Wishart language, corresponds to line matrices $Z$ of size $N\times 1$. 

If $(z_\alpha)_{\alpha\in \llbracket 1,N\rrbracket}\in \mathbb{C}^N$, then the matrix
\[ \rho_{\alpha\beta}:= \frac{z_\alpha\overline{z}_\beta}{\sum_{\gamma \in \llbracket 1,N\rrbracket} z_\gamma\overline{z}_\gamma}\]
is a rank one density matrix and every such matrix arises in that way. Moreover, any unitary invariant measure on $\mathbb{C}^N$ gives rise to a unitary invariant measure on rank one density matrices. There is actually only one such measure: the space of rank one density matrices on is $\mathbb{C}^N$ is simply the projective space $\mathbb{P}(\mathbb{C}^N)$, on which the unitary group acts transitively. One can work with a simple measure on  $\mathbb{C}^N$, namely
\beq \label{eq:Gauss-z}
 \prod_{\alpha\in \llbracket 1,N\rrbracket} \frac{d^2z_\alpha}{\pi} e^{-z_\alpha\overline{z}_\alpha},
 \eeq
which is the only one such that the measure is unitary invariant and the components are independent. But any normalised measure
$\prod_{\alpha\in \llbracket 1,N\rrbracket} d^2z_\alpha f(||z||)$, with $||z||^2=\sum_{\gamma \in \llbracket 1,N\rrbracket} z_\gamma\overline{z}_\gamma$, 
induces the same measure on the space of rank one density matrices (as a simple consequence of the homogeneity of the formula for $\rho$ in terms of the $z$'s). Uniqueness is clear in the formula \eqref{eq:HCIZ} or \eqref{eq:HCIZ-N}.

In the large $N$ limit, the expression for $N\rho$ can be replaced by the simpler formula,
\[ N \rho_{\alpha\beta}\simeq  z_\alpha\overline{z}_\beta, \]
if the Gaussian measure \eqref{eq:Gauss-z} on $(z_\alpha)_{\alpha\in \llbracket 1,N\rrbracket}$ is used. Indeed, the random variables  $z_\gamma\overline{z}_\gamma$ are easily seen to be independent exponential (do the angular integral) so  $\frac{1}{N}\sum_{\gamma \in \llbracket 1,N\rrbracket} z_\gamma\overline{z}_\gamma$ concentrates near the value $1$, up to fluctuations of order $O(\frac{1}{\sqrt{N}})$. It is clear that, within this approximation, $N \rho_{\alpha\beta}=  z_\alpha\overline{z}_\beta$, for $N$ fixed, and
\beq
 \mathbb{E}(e^{N\text{Tr} \, A\rho})= \int \prod_{\alpha\in \llbracket 1,N\rrbracket} \frac{d^2z_\alpha}{\pi} e^{-z_\alpha\overline{z}_\alpha} e^{\sum_{\alpha,\beta\in \llbracket 1,N\rrbracket}A_{\alpha\beta} z_\alpha\overline{z}_\beta }=\frac{1}{\text{Det}\,(1-A)},
 \eeq
by a standard Gaussian integral formula valid for hermitian matrices $A$  with no eigenvalue $>1$. One infers that if $N$ gets large but $A$ satisifies certain conditions, in particular if, starting from a certain size, say $n$,  $A_{\alpha \beta}$ evolves with $N$ by completing with zeros, i.e. $A=\begin{pmatrix} A^{(nn)} & 0^{n(N-n)}  \\ 0^{(N-n)n} & 0^{(N-n)(N-n)}  \end{pmatrix}$, then 
\beq
 \lim_{N\to \infty} \mathbb{E}(e^{N\text{Tr} \, A\rho})= \lim_{N\to \infty} \int \prod_{\alpha\in \llbracket 1,N\rrbracket} \frac{d^2z_\alpha}{\pi} e^{-z_\alpha\overline{z}_\alpha} e^{\sum_{\alpha,\beta\in \llbracket 1,N\rrbracket}A_{\alpha\beta} z_\alpha\overline{z}_\beta }=\frac{1}{\text{Det}\,(1-A)}.
 \eeq
This coincides with \eqref{eq:Nrho-A}. It gives an explicit representation of the limiting object ``$\lim_{N\to \infty} N \rho$''.

More generally, in the large $N$ limit with a subsystem of size $M$, we may split the indices and consider $A_{\alpha i, \beta j}=a_{\alpha\beta}\delta_{ij}$ with random variables $z_{\alpha i}$, with $\alpha \in \llbracket 1,M\rrbracket $, $i\in \llbracket 1,\tilde{M}\rrbracket$ with $M\tilde{M}=N$. The random variable $\rho^{red}$ with matrix elements, 
\beq \label{eq:Mred-rho}
 \rho^{red}_{\alpha\beta} :=\frac{\sum_i z_{\alpha i}\overline{z}_{\beta i}}{\sum_{\gamma i} z_{\gamma i}\overline{z}_{\gamma i} },
 \eeq
which couples to $a=(a_{\alpha\beta})$ has then a simple limit distribution (as we shall see). We can alternatively write the reduced density matrix as $\rho^{red}= \frac{Z Z^\dag}{\Tr ZZ^\dag}$ with $Z$ the complex rectangular matrix with entries $(z_{\alpha i})$. The measure on $Z$ is then that of Wishart ensembles \eqref{eq:wishart}.

We first investigate the large $\tilde{M}=\frac{N}{M}$ limit, $M$ fixed. Let us define $\xi_{\alpha\beta}^{({M})}$ by
\beq \label{eq:def-xiM}
 \sum_i z_{\alpha i}\overline{z}_{\beta i}=: N\Big( \frac{\delta_{\alpha\beta}}{M}+  \frac{1}{\sqrt{NM}} \xi_{\alpha\beta}^{({M})}\Big),
 \eeq
so that
\beq \label{eq:def-xiM}
 \rho^{red}_{\alpha\beta}=\frac{N \delta_{\alpha\beta}+\sqrt{NM}\xi_{\alpha\beta}^{({M})}}{NM+ \sqrt{NM}\sum_{\gamma}\xi^{({M})}_{\gamma\gamma}}.
 \eeq
It is easy to check that if $a=(a_{\alpha\beta})$ and $\xi^{({M})}=(\xi^{({M})}_{\alpha\beta})$ then, at $M$ fixed,
\beq
 \lim_{N\to \infty} \mathbb{E}(e^{\text{Tr} \, a\xi^{({M})}})=e^{\frac{1}{2}\text{Tr} \, a^2}.
 \eeq
Thus, as expected from the central limit theorem,  $\xi^{({M})}$ converges in law at large $N$ to an hermitian Gaussian random matrix (with the simplest covariance). We denote this limit (in law) by $\xi=(\xi_{\alpha\beta})$ so that $\mathbb{E}(e^{\text{Tr} \, a\xi})=e^{\frac{1}{2}\text{Tr} \, a^2}$.

We can now translate this result as a statement on the reduced density matrix \eqref{eq:Mred-rho}. Using the definition \eqref{eq:Mred-rho} and the representation \eqref{eq:def-xiM}, it is easy to verify that, at large $N$, M fixed, we get
\[ \sqrt{NM} \left(\rho^{red}_{\alpha\beta} -\frac{\delta_{\alpha\beta}}{M}\right) \stackrel{\text{law}}{\rightarrow} \xi_{\alpha\beta}-\frac{\delta_{\alpha\beta}}{M}\sum_{\gamma}\xi_{\gamma\gamma} . \]
This coincides with \eqref{eq:Zgauss}. It gives a an explicit representation of the limiting object ``$\lim_{N\to \infty} \sqrt{N} \left(\rho^{red}_{\alpha\beta} -\frac{\delta_{\alpha\beta}}{M}\right)$''.

\end{document}